Blockchain technology for a Safe and Transparent Covid-19 Vaccination

Maha FILALI ROTBI[1], *, Saad MOTAHHIR[2], Abdelaziz EL GHZIZAL[1]

[1] Innovative Technologies Laboratory, EST, SMBA University, Fez, Morocco
[2] Engineering, Systems and Applications Laboratory, ENSA, SMBA University, Fez, Morocco

*: maha.filalirotbi@usmba.ac.ma

**Abstract**

In late 2019, we witnessed the apparition of the covid-19 virus. The virus appeared first in Wuhan, and due to people travel was spread worldwide. Exponential spread as well as high mortality rates, the two characteristics of the SARS-CoV-2 virus that pushed the entire world into a global lock-down. Health and economic crisis, along with social distancing have put the globe in a highly challenging situation. Unprecedented pressure on the health care system exposed many loopholes not only in this industry but many other sectors, which resulted in a set of new challenges that researchers and scientists among others must face. In all these circumstances, we could attend, in a surprisingly short amount of time, the creation of multiple vaccine candidates. The vaccines were clinically tested and approved, which brought us to the phase of vaccination. Safety, security, transparency, and traceability are highly required in this context. As a contribution to assure an efficient vaccination campaign, in this paper we suggest a Blockchain-based system to manage the registration, storage, and distribution of the vaccines.

## 1. Introduction

Declared by the World Health Organization (WHO) on January 30, 2020, as a global pandemic [21]. An analysis of the Covid-19 spread showed that most countries endure an exponential raise of Covid-19 infection cases [22]. As a response to lighten the easy spread of the virus, the countries took total lockdown as a measure to control its propagation, which led to rising levels of physical and mental domestic violence among many other forms of social abuse [23].
The covid-19 virus caused a significant Socioeconomic crisis. An analysis of the impacted sectors was presented in [24].
In the context of combating the coronavirus crisis, a blockchain and AI-based architecture was introduced in [20]. The authors of this paper proposed five key blockchain-based solutions to address the following points: outbreak tracking, user safety, daily operations, medical supply chain, and donation tracking.
The news of the first successful vaccine trials has the potential to relieve billions of people all over the world. However, as encouraging as these trials were, it is critical to remember that a vaccine cannot be as effective if it is not correctly distributed and trusted by the public.
Vaccine production is an endeavor in which an almost infinite number of combinations must work flawlessly, on the other hand, the management of the produced vaccine lots, and their distribution is a whole other process that has to be cautiously handled to prevent the loss of effectiveness.
Blockchain and IoT-based platforms can provide a structure to assure the effectiveness of the vaccination operation. BIoT improves system performances and simplifies decision-making for businesses. Such systems can aid in the development of trust, the reduction of costs, and the acceleration of transactions. Furthermore, decentralization, which is at the heart of blockchain technology, has the potential to eliminate single points of failure in an IoT network [6].

For this vaccination to be successful, the following aspects need to be considered: data correctness, data immutability, data traceability, transparency, security, and all the final that is beneficiaries' safety. A blockchain-based system for covid-19 vaccination was presented, which handles patients' registration and self-reporting on one hand, on the other hand, it allows the registration of vaccine, storage, and distribution monitoring [9].

In [19], A general Blockchain model design was suggested to monitor the different phases of the vaccine supply chain.

In a recent study, IBM collaborated with organizations at various stages of the pharmaceutical supply chain to demonstrate how data in a blockchain network can potentially be used for covid-19 vaccination [18]. IBM company has also recently developed a Blockchain-powered digital health passport platform to allow individuals to store and share their health status while protecting their privacy. Moreover, individuals can also present their health information through their health pass rather than sharing personal or sensitive information, such as lab results or medical history, with a large group of people. The platform allows individuals to control who can access their data, as a result, each individual must consent to the release of their personal health data as part of their health status. Cathay Pacific Airways and United Airlines are currently conducting two trials. According to the press release, the trials will attempt to replicate the full traveler experience from taking a COVID-19 test before departing to uploading the results and then following entry requirements at their departure and destination airports.[16].

The Government of Singapore has HealthCerts - a set of digital open standards and schema for issuing digital COVID-19 test results/vaccination certificates [14]. To boost downstream authentication, digital certificates can be digitally authenticated and endorsed by the government via Notarise.. Verify provides an easy, reliable way to ensure digital certificates have not been tampered with and issued/digitally authenticated and endorsed by a recognized entity.

An analysis of 15 marketplaces revealed that three main Covid-19 vaccines are being advertised for sale on the darknet, the hidden part of the internet that requires specialized software to access[17]. According to the study, Covid-19 products are available in 12 markets, with three markets accounting for 85 percent of all 645 listings [17].

Even though the majority of vaccines on the dark-net are likely to be scams but among the fake advertisements, there will be some genuine product delivery of sellers who have supply chain connections or access to leftover product or product that has fallen off the back of a truck.

In this paper, we present an approach for a BIoT system to support the covid-19 vaccination, as well as other vaccinations. The system identifies doses, manages patients' registrations, as well as vaccine registrations. Temperature excursions are handled using IoT along with smart contracts. This work aims to assure the safety and effectiveness of vaccine doses before they reach patients. The BIoT system provides transparency, the correctness of data, data immutability, and decentralization.

The article is organized as follows, first, we present some of the main covid-19 vaccines and their characteristics, also the parameters to be monitored to protect the vaccine vials, next we discuss the reasons why blockchain is suitable for this application, later we present the main actors of the platform and their interactions, then we discuss the functionalities handled using smart contracts. In chapter 4, we discuss how the system can be used post-vaccination. Challenges are presented in chapter5. Chapter 6 presents the Conclusions.

## 2. Covid-19 vaccine:

The apparition of the virus attracted attention first in Wuhan, China in late December 2019, after that by early January 2020 the contagious virus was referred to as "Wuhan coronavirus", but by February 2020, it was officially announced as "severe acute respiratory syndrome coronavirus

2"(SARS-CoV-2) later it was officially renamed by the WHO as Covid-19 [3]. As the coronavirus kept spreading at an exponential rate, many industries worldwide including the pharmaceutical and health sectors were facing a challenging situation. Health care specialists were aware that the traditional process of vaccine development wasn't possible in this case, due to the potential spread of the virus and the mortality rate. The rapid rollout of the virus required modern, innovative and rapid solutions. The development process of vaccines usually requires decades so that the vaccine can be approved, the availability of many vaccine candidates against the covid-19 virus before the end of 2020, is considered uncommon and unprecedented. Thanks to the new manufacturing platforms, protein engineering, computational biology, and gene synthesis provided the necessary tools and made it possible to witness the development of various vaccines with the speed and precision needed to answer the global need [1].

Most of the vaccines that were developed against covid-19, use copies of the same protein that exists on the surface of the covid-19 virus, their function is to train the immune system so it will be able to recognize it, in case of an ulterior infection.

The mRNA vaccine, for instance, leads the human cells to produce the same spike protein of the virus. The downside with these vaccines is the fact that they are fragile and become useless if not preserved at ultra-low temperatures [4]. With that being said, the control and monitoring of such parameters that impact the efficiency and the safety of the vaccine supply is a universal priority. Hereafter a list of some of the parameters and data that can be monitored and registered during the life cycle of the vaccines to prevent the loss of their effectiveness and keep track of the historical values of these parameters for ulterior use.

- Initial set-up and Freezer storage units

Temperature: Maximum, minimum, and current temperature need to be recorded
  Exposure to light
- Point of distribution
  Time of distribution
  Current, Maximum, and minimum temperatures of all the units
  Exposure to light
- Transportation
  Temperature
  Humidity
  Transportation duration
- Receipt of vaccine

Date and time of receipt of the vaccines
  Temperature
  Storage duration

Following a table that presents two types of covid-19 vaccine and some properties and parameters ranges to be cautiously controlled to protect the effectiveness of the vaccines [4], [2], [12].

| Storage Condition | Pfizer-BioNTech | Moderna | Covaxin by Bharat Biotech | Covishield by Serum Institute |
|---|---|---|---|---|
| Frozen Vials Prior to use | Temperature between -80°C and -60°C Protection from light | Temperature between -25°C and -15°C Protection from light Avoid storage on dry ice or | A vaccine with no sub-zero storage, no reconstitution requirement, and ready to use liquid | Do not freeze. Store in a refrigerator (+2oC to +8oC). |

|  | Recommended storage temperature is -70°C | below -40°C Recommend storage temperature is – 20°C Shelf life is 6 months + 30 days further at refrigerated state | presentation in multi-dose vials, stable at 2-8°C. |  |
| --- | --- | --- | --- | --- |
| Thawed, unpunctured vials | The vaccine can be thawed and stored at temperatures ranging from +2°C to +8°C for up to 120 hours or at room temperature (up to +25°C ) for no more than 2 hours. | If the vaccine has not been punctured, it should be thawed and stored for up to 30 days at +2°C to +8°C, or for up to 12 hours at +8°C to +25°C. Avoid exposure to light during storage | Storage temperature should be between 2 to 8°C. | Storage temperature should be between 2 to 8°C. |
| Thawed, punctured vials | After dilution, storage temperature should be between +2°C to +25°C Minimize exposure to light during storage | Storage temperature should be between +2°C to below +25°C Vials should be protected from light | Storage temperature should be between 2 to 8°C. | Once opened, multi-dose vials should be used as soon as practically possible and within 6 hours when kept between +2°C and +25°C |

## 3. Blockchain technology for Covid-19 vaccine:

Vaccines are fragile biological substances, and if exposed to temperatures (heat or cold) outside of the product's recommended range (i.e., ultra-low or frozen temperatures) or opened to non-tolerated light can lose their efficacy and effectiveness.
Storage, distribution, and administration of vaccines is a very complex process that needs cautious handling.
When it comes to vaccines, beneficiary safety is the most crucial point. In this paper, we discuss how blockchain technology can help to automate the vaccine lots' management process, in a way to simplify the operation, make it transparent to the different members, and secure.
In the following subsections, we give a quick understanding of Blockchain technology, how can any vaccination campaign, and especially the covid-19 vaccination, benefit from this technology, later we will present the main actors of the proposed system, and their interaction with it.

## 3.1. Blockchain technology:

Blockchain can be defined as a tamper-resistant distributed ledger, where no central authority has control over information. It operates in a peer-to-peer network where all nodes interact with each other. Blockchain at its basic level enables nodes belonging to the same network to record transactions in the shared ledger of the network.

The invention of Bitcoin in 2008 exposed the world to a new concept that revolutionized society as a whole. It was something that had the potential to influence any industry. Blockchain was this new notion.

As mentioned earlier, blockchain was first introduced in the context of cryptocurrencies, later in 2013, researchers and developers noticed the potential of this technology and its promising application in other areas.

Being independently managed and maintained by a distributed network of peers, along with cryptographic mechanisms, makes blockchain secure and resilient to attempts to alter data stored in the digital ledger [5].

The graph in Figure 1 depicts a broad overview of the year-by-year evolution and adoption trends of blockchain technology [6].

Interest in blockchain has exponentially risen during the past few years, the largest companies and organizations around the world are investing millions of dollars to experiment and implement this technology in their projects. Various blockchain technology applications, including Government, Energy, Smart Cities, Transportation, Healthcare…, are mentioned in [7].

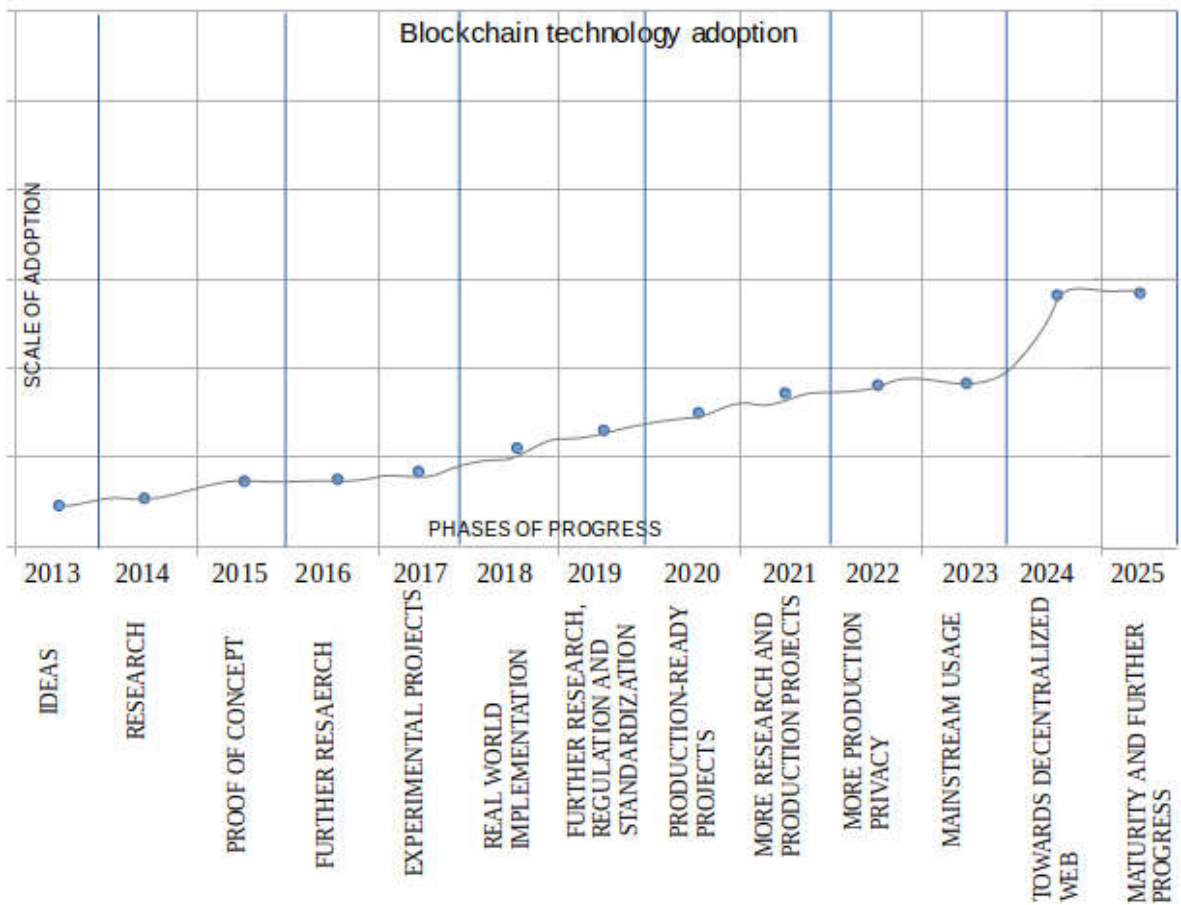

*Figure 1: Adoption of Blockchain Technology*

In our paper, we present a blockchain technology application in the healthcare industry, and more precisely for the Covid-19 vaccination campaign.
Before we dive into subsections about why and how Blockchain can assist the Covid-19 vaccination, we will briefly present some blockchain terminology to help understand the coming sections.

### 3.1.1. Blockchain Terminology:

Peer-to-Peer Network: A network topology that allows peers to communicate with each other directly without any centralized server.
Transaction: a transfer of value between two users.
Block: composed of a number of transactions and other elements (timestamp, hash pointer…).
Hash pointer: hash of the previous block.
Smart Contract: a computer program that runs on top of the blockchain and is executed when a set of conditions are satisfied.
Hash Function: an irreversible function that takes an input value and turns it into a unique hexadecimal number. For the same input value, we always have the same output.

### 3.2. Reasons why Blockchain is Suitable for Covid-19 Vaccination:

Blockchain isn't necessarily the best solution for any digital platform. It's not evident to choose between a centralized ledger and a blockchain. In [8], the authors provided a structured methodology to determine whether blockchain technology is suitable for a given scenario or not based on the application requirements.
In the context of our application, we present a list of the main features that blockchain offers, and that meets our need:
- Many participants: more than one actor participates in the vaccination process.
- Distributed participants
- Lack of trusted third party: no centralized access and control of data
- Data monitoring in real-time: Vaccines parameters should be monitored in real-time.
- Full transactional history shared between participants: data recorded to the ledger is shared to guarantee the transparency of the vaccination.

To determine if blockchain is appropriate in our case, figure 2 shows a flowchart of the steps to follow.

### 3.3. Blockchain-based platform for Covid-19 vaccination:

As described in [10], blockchain technology offers the necessary features and structure to manage the vaccination campaign from pharmaceuticals to patient arms.
Blockchain offers a transparent record system shared among all the participants of the supply chain. To protect the vaccine vials and alert in case of the nonconformity of values, vaccine temperature outside the recommended range, smart contracts are used to verify the correctness and conformity of collected data.
Blockchain provides immutability and independent verification of data records, which makes the vaccination process transparent.

In the following subsections, we present more details about the main actors, the architecture of the platform, and the use of smart contracts for secure vaccine management.

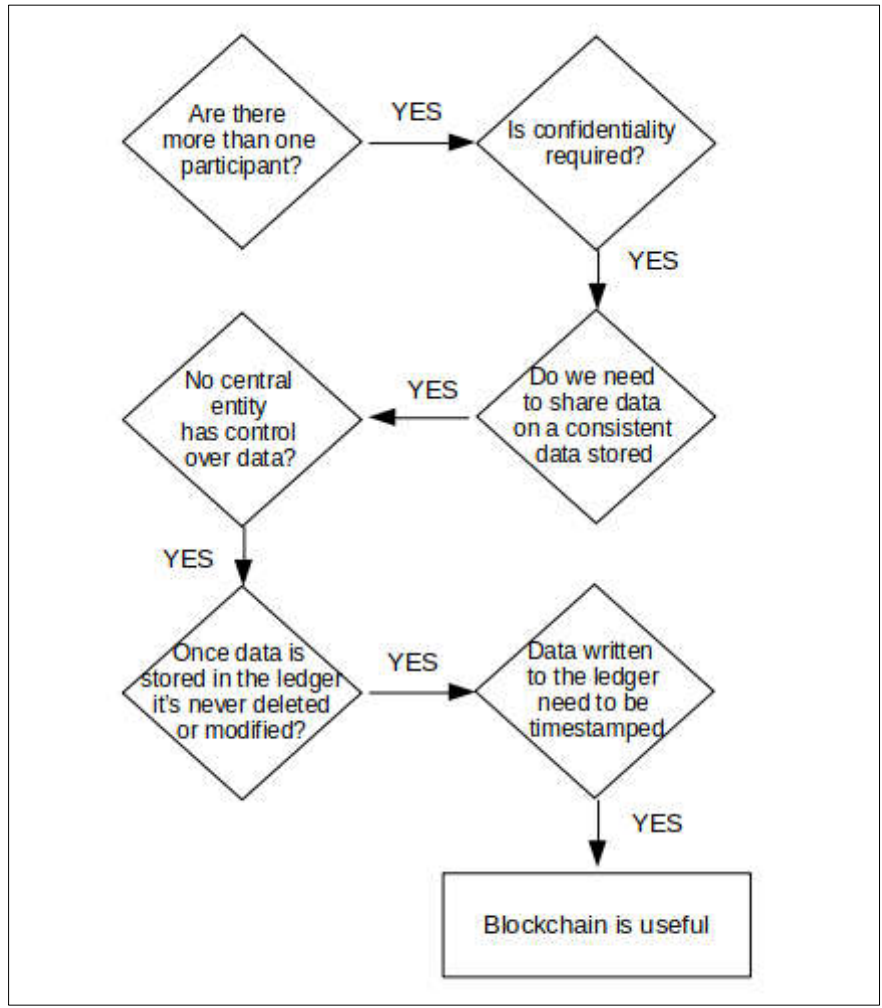

*Figure 2: Blockchain usage flowchart*

### 3.3.1. Main actors of the platform

The main actors that contribute to the Blockchain-based system for covid-19 vaccination are:
- Vaccine Manufacturer
- Vaccine distributor
- Beneficiaries/Patients
- Doctors
- Medical center

Many doubts about the offered vaccines have risen. Some people hesitate and even refuse to get vaccinated for the simple reason that they are concerned about their health and don't trust the untransparent vaccination process. Reporting side effects to the platform is one of the features that can help patients build trust in the vaccination campaign by consulting the beneficiaries' reported experience directly.

In the platform we suggest, the interaction of the participants with the application is as shown below

| Vaccine Manufacturer | Vaccine Distributor |
|---|---|
| Interaction with the platform: Prepares, identifies, and registers the vaccine lots to the system. Defines rules and conditions of vaccine storage and distribution. Records transportation date and time to the system | Interaction with the platform: Scans the vaccine lot ID. Verifies the vaccines state in the system before transporting the lots. Links the Vaccine lot ID to the Vehicle ID |

| Medical center | Doctor |
|---|---|
| Interaction with the platform: Verifies the vaccines state in the system. Verifies the match (Vaccine ID, Vehicle ID) by scanning both IDs. Records reception date and time to the system. Records time of vaccine storage in the storage unit | Interaction with the platform: Registers the patient for Vaccination. Verify the Beneficiary ID. Scans the vaccine ID for verification before giving the vaccine shot. Associate BID with VID |

| Patient/Beneficiary |
|---|
| Interaction with the platform: Registers to the system. Scans the vaccine ID for verification before getting the vaccine shot. **Reports side effects** |

### 3.3.2. BIoT Application architecture

A Blockchain-based Internet of Things architecture, presented in [6], is used as the architecture for vaccination management.
This architecture is composed of 6 layers:
1. **Physical layer**: includes real-world objects of the application that need to be monitored and controlled.
2. **Device layer:** contains devices such as sensors, actuators, RFID tags...
3. **Network layer**: offers connectivity between devices.
4. **Blockchain layer**: a middleware between participants of the network.
5. **Management layer**: includes platforms that allow data collected from IoT devices to be processed.
6. **Application layer**: the final application.

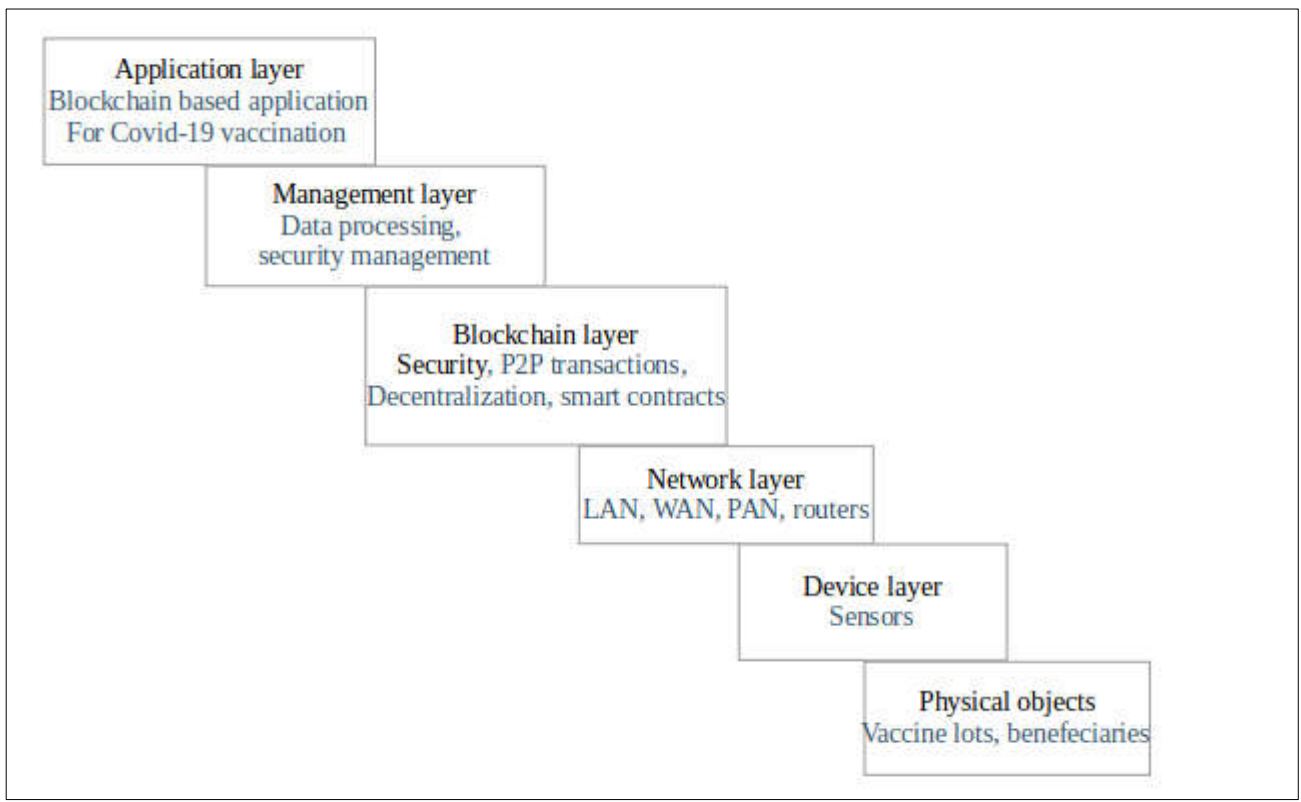

*Figure 3: BIoT Architecture*

### 3.3.3. Smart contracts:

Blockchain enables things to communicate directly with each other, and with the availability of smart contracts, negotiations and all types of transactions and contracts can also occur directly between the participants of the system without requiring an intermediary or an authority.
Smart contracts can be used throughout the cold chain to ensure that desirable conditions are maintained during storage and transportation, as well as to issue warning signs if sensors detect any abnormalities in the handling of vaccines.
Following are the functionalities to be implemented using smart contracts:

- Associating the transportation vehicle identifier (TID) to the vaccine lot identifier (VID)
- Associating beneficiary identifier (BID) with VID
- Automates the vaccination operation dates, based on the received patients' registration requests (BRR) and the available vaccine vials.
- Assuring on-time delivery vials based on BRR, locally available vials, and transportation time.
- Alert feature in case temperature collected values are outside the recommended range
- Reported side effects management
- Reported adverse events management
- Providing vaccination certificate

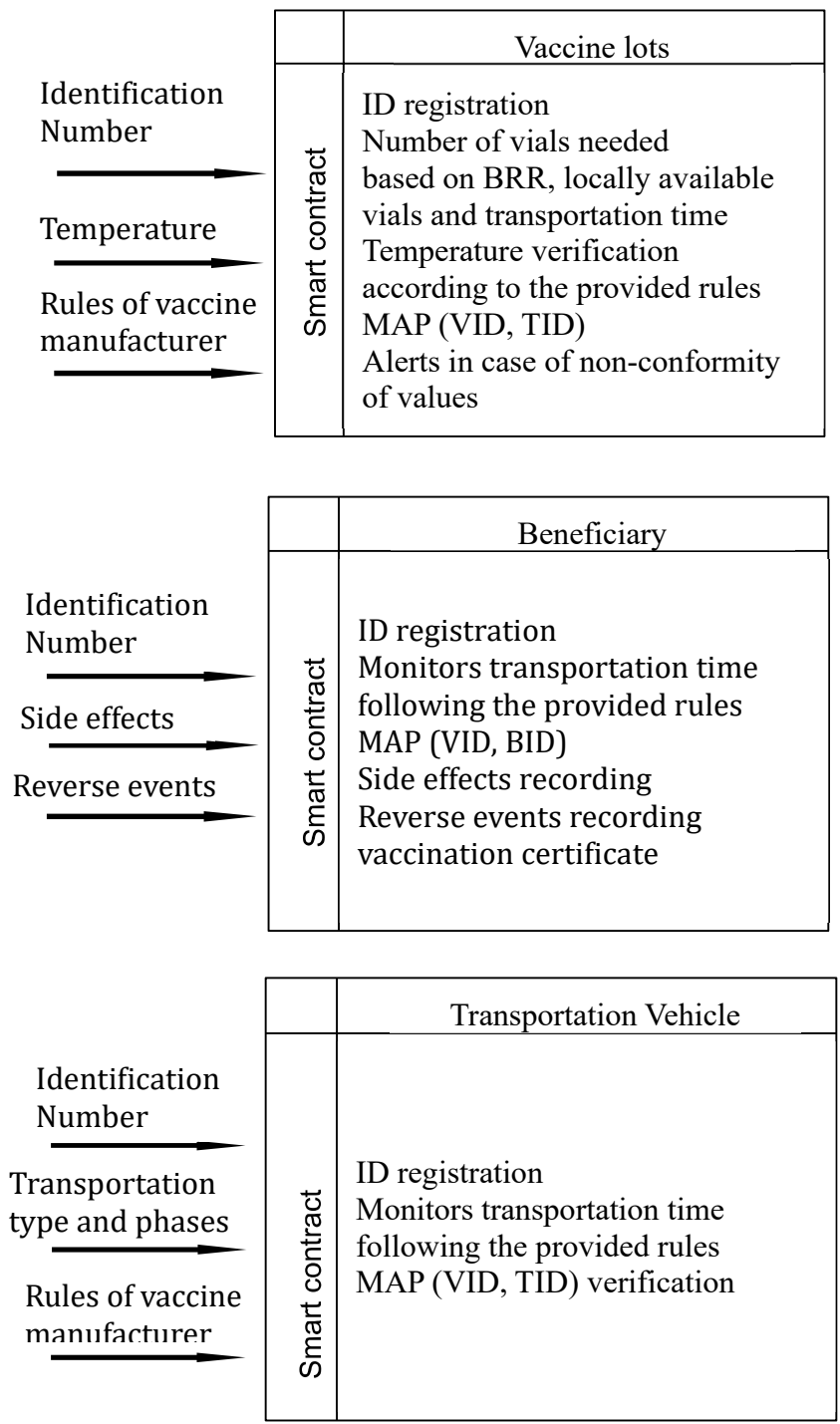

For more information about implementing blockchain in the supply chain, logistics and transport the authors in [11] presented a review of many works related to this area.

**3.3.4. Implementation of the suggested platform:**

For the implementation of this solution, we opt for the Ethereum 2.0 blockchain technology. The development of our smart contracts can be done using solidity programming language.
Ethereum blockchain allows developers to build applications that take advantage of blockchains' decentralization and security features, while eliminating the need to create a new blockchain for each new application. Ethereum can be thought of as "the smartphone" of blockchain, it provides the ecosystem to build personalized applications [25].
In our case Ethereum provides process authentication, furthermore verification and analysis of transactions through mining operation.
Solidity is a high-level programming language, it supports inheritance and polymorphism, also libraries, and complex user-defined types. It is a Turing-complete language that has JavaScript-like syntax. When using Solidity, contract design is similar to classes in object-oriented programming languages [26].

**4. Covid-19 vaccine as a start of a digitalized health care system:**

Implementing such a system for covid-19 vaccination makes post-covid management easier. All the countries are now working on administering the vaccine to their citizens.
Currently, besides selling vaccine doses, people on the darknet are even giving vaccination records to help their customers travel easily [17]. The need for secure and consistent management of beneficiaries and a fully digitalized system is highly required.
There are huge discussions on how Countries can give their citizens Vaccine Certificates to open cross-border economic activity and travel. The need is to provide a Digital Certificate which aligns with the WHO, Global Standards, and individual country Government requirements.
Part of the suggested platform includes vaccination certification. Once the beneficiary gets the first vaccination, the application allows the certificate delivery as a QR code. The QR code contains the patient's ID, the vaccine type, and vaccination date. A final certificate is offered after the validation of the second shot. The suggested platform is profitable in the case of vaccinations other than covid-19 as long as all the participants contribute with the needed data and rules. It's also beneficial in case of adverse events – Any health problem that may occur after a shot or other vaccine might appear later.
Even after the pandemic is under control, we need a system of securely sending and receiving Health Records and maintaining an immutable ledger to store data privately. We can start digitalizing the health records since the covid-19 vaccination, or since the birth of an individual, and help citizens use it wherever they would like to. The suggested system can provide a bridge to a distributed, immutable, and decentralized health care system.

**5. Open challenges:**

One of the major challenges we face is the authenticity and ethical behavior of the system participants. In the case where storage unit malfunction is not addressed, or electricity disruption is not fixed, might lead to ineffective vials of Vaccine.
In cases where IoT devices cannot be used, we need to be at the mercy of the actor's ethical behavior.

One more major challenge is keeping the Smart Contracts immutable and their presence in a device might lead to any party altering it and making it a private blockchain instead of a Public blockchain. All the smart contracts must follow permissioned login and all transactions must be declared to all participants making it a truly peer-to-peer network with highly immutable smart contracts.
Another challenge is our need for participants in the supply chain to set up the environment and their commercial agreements.
The final concern is from energy conservationists about the Carbon emissions of Blockchain Mining. Cambridge's Center for Alternative Finances estimates that bitcoin's annualized electricity consumption hovers just above 115 terawatt-hours (TWh) while
Digiconomist's closely tracked index puts it closer to 80 Twh[13]. A single Ethereum transaction absorbs more energy than an average U.S. transaction. uses in a day[14].
There are various ways in which this matter can be solved for example using Proof of Stake(PoS) instead of Proof of Work(PoW) consensus algorithm. This reduces the energy consumptions and hence the lowered amount of carbon footprints. We have to make sure we keep a balance of energy consumption as well as data in the blockchain safe and secured.

6. Conclusions:

Challenging situations initiate innovative solutions, which aligns with the context of this work. In this paper, we present a system to serve the covid-19 vaccination campaign based on blockchain technology, along with the internet of things.
Our system encapsulates the whole vaccination process, citizen registration, vaccine storage monitoring, vaccine transportation monitoring, vaccination organization, side effects, adverse events reporting, and Post-Vaccination management. The suggested BIoT architecture offers a transparent, secure, and immutable platform that allows the verification of vaccines lots state, which is represented by the parameters collected by the Device layer, and whether or not they respect the rules and conditions set by the vaccine manufacturer. Worth mentioning that the platform suggested can be used for any vaccination operation.